\documentclass[twocolumn]{aastex6}
\usepackage{graphicx,epsfig,amssymb,amsmath,color,lineno}

\def\eg{{e.g.,~}}
\def\ie{{i.e.,~}}

\newcommand{\p}{^\prime}
\newcommand{\e}{\epsilon}

\newcommand{\ep}{\epsilon^\prime}

\newcommand{\psim}{\lower.5ex\hbox{$\; \buildrel \propto \over\sim \;$}}
\newcommand{\lbar}{\lower.0ex\hbox{$\; \buildrel
{\lower0.0ex \hbox{-}} \over\lambda  \;$}}

\slugcomment{Draft; \today}

\shorttitle{$H_{0}$ and $\Omega_{m}$ using $\gamma$-ray attenuation}
\shortauthors{Authors}

\begin{document}
\label{firstpage}

\title{A new measurement of the Hubble constant and matter content of the Universe using extragalactic background light $\gamma$-ray attenuation}
\author{{\sc A. Dom\'inguez}\altaffilmark{1}, {\sc R. Wojtak}\altaffilmark{2}, {\sc J. Finke}\altaffilmark{3}, {\sc M. Ajello}\altaffilmark{4}, {\sc K. Helgason}\altaffilmark{5}, {\sc F. Prada}\altaffilmark{6}, {\sc A. Desai}\altaffilmark{4}, {\sc V. Paliya}\altaffilmark{7}, {\sc L. Marcotulli}\altaffilmark{4}, {\sc D.~H. Hartmann}\altaffilmark{4}}
\altaffiltext{1}{IPARCOS and Department of EMFTEL, Universidad Complutense de Madrid, E-28040 Madrid, Spain; alberto.d@ucm.es}
\altaffiltext{2}{DARK, Niels Bohr Institute, University of Copenhagen, Lyngbyvej 2, 2100 Copenhagen, Denmark; radek.wojtak@nbi.ku.dk}
\altaffiltext{3}{U.S. Naval Research Laboratory, Code 7653, 4555 Overlook Avenue SW, Washington, DC, 20375-5352}
\altaffiltext{4}{Department of Physics and Astronomy, Clemson University, Kinard Lab of Physics, Clemson, SC 29634-0978, USA}
\altaffiltext{5}{Science Institute, University of Iceland, IS-107 Reykjavik, Iceland}
\altaffiltext{6}{Instituto de Astrofísica de Andaluc{\'i}a (CSIC), Glorieta de la Astronom{\'i}a, E-18080 Granada, Spain}
\altaffiltext{7}{Deutsches Elektronen-Synchrotron, D-15738 Zeuthen, Germany}


\begin{abstract}
The Hubble constant $H_{0}$ and matter density $\Omega_{m}$ of the Universe are measured using the latest $\gamma$-ray attenuation results from {\it Fermi}-LAT and Cherenkov telescopes. This methodology is based upon the fact that the extragalactic background light supplies opacity for very high energy photons via photon-photon interaction. The amount of $\gamma$-ray attenuation along the line of sight depends on the expansion rate and matter content of the Universe. This novel strategy results in a value of $H_{0}=67.4_{-6.2}^{+6.0}$~km~s$^{-1}$~Mpc$^{-1}$ and $\Omega_{m}=0.14_{-0.07}^{+0.06}$. These estimates are independent and complementary to those based on the distance ladder, cosmic microwave background (CMB), clustering with weak lensing, and strong lensing data. We also produce a joint likelihood analysis of our results from $\gamma$ rays and these from more mature methodologies, excluding the CMB, yielding a combined value of $H_{0}=66.6\pm 1.6$~km~s$^{-1}$~Mpc$^{-1}$ and $\Omega_{m}=0.29\pm 0.02$.

\end{abstract}


\section{Introduction}
\label{sec:intro}
Very high energy photons (VHE, $E\geq 30$~GeV) travel impeded through the Universe due to the extragalactic background light (EBL). The EBL is a diffuse radiation field that fills the Universe from ultraviolet through infrared wavelengths, and is mainly produced by star formation processes over cosmic history \citep[\eg][]{hauser01}. A $\gamma$-ray and an EBL photon may annihilate and produce an electron$-$positron pair \citep{nikishov62,gould66}. This interaction process generates an attenuation in the spectra of $\gamma$-ray sources above a critical energy. This effect may be characterized by an optical depth and has been observed with the current generation of $\gamma$-ray telescopes \citep[\eg][]{ebl12,hess_ebl13,dominguez13a,biteau15,dominguez15}. The proper density of the absorbing EBL photons along the line of sight depends on the expansion history and is therefore cosmology dependent. This fact allowed \citet{dominguez13b} to measure the local expansion rate of the universe, \ie the Hubble constant $H_0$. Their $H_{0}$ measurement is based upon a comparison between an observational estimate of the cosmic $\gamma$-ray horizon \citep[CGRH\footnote{The CGRH is defined as the energy at which the EBL absorption optical depth is equal to unity as a function of redshift \citep{fazio70}.}, ][]{dominguez13a} and those derived from an empirical EBL model \citep{dominguez11a}.

Recently, the {\it Fermi}-LAT collaboration has published an unprecedented measurement of the optical depths as a function of energy in the redshift range that goes from the local Universe to a redshift of approximately 3 \citep{ebl18}. Furthermore, \cite{desai19} have provided complementary optical depth measurements at higher energies than \citet{ebl18}, focused in the lower redshift Universe. These latter results are based on blazar observations with Imaging Atmospheric Cherenkov telescopes (IACTs). In the present analysis, we take advantage of these recent products by comparing them with optical depth estimates based on the EBL models developed by \citet{finke10} and \citet{dominguez11a}. Our methodology allows us to use $\gamma$-ray absorption to constrain $H_0$ and $\Omega_{m}$ simultaneously for the first time.  We include an estimate of the systematic uncertainties, including those by the EBL models.

As discussed by \citet{suyu12}, multiple paths to independent determinations of the Hubble constant are needed in order to assess and control systematic uncertainties \citep[see also,][]{chen11,abbott17,des19,freedman19}. Accurate estimates of $H_{0}$ provide critical independent constraints on dark energy, spatial curvature, neutrino physics, and general relativity \citep{freedman10,suyu12,weinberg13}. Moreover, independent determinations of the Hubble constant will play a crucial role in resolving the problem of a discrepancy between the Hubble constant inferred from the cosmic microwave background radiation (CMB) and those from type Ia supernovae with distance calibration from Cepheids \citep{riess18a}. This tension may likely result from currently unknown systematic effects, but it can also signify an intrinsic inconsistency within the standard $\Lambda$CDM \citep[e.g.~][]{val2016,Woj2017}.

Here, we derive a complementary estimate of $H_{0}$ and $\Omega_{m}$ using EBL attenuation data. This paper is organized as follows. Section~2 gives theoretical and observational backgrounds, whereas Section~3 shows general considerations and describes our likelihood methodology. In Section~4, we present the cosmological results. Finally, a discussion and summary of our results is presented in Section~5.

\section{Theoretical and observational background} \label{sec:background}

The potential of measuring the Hubble constant from $\gamma$-ray attenuation was already pointed out two decades ago by \citet{salamon94} and \citet{mannheim96}, when the $\gamma$-ray experiments at that time could only study a few sources on the entire sky. Later, \citet{blanch05a,blanch05b,blanch05c} studied, in a series of papers, the potential of using the CGRH to constrain cosmology. These investigations were motivated by the starting operation of IACTs such as MAGIC, VERITAS, and H.E.S.S. \citep[][respectively]{lorenz04,weekes02,hinton04}. Blanch \& Mart\'inez used simulated VHE spectra of blazars, at different redshifts, to estimate how some relevant cosmological parameters could be constrained. Their analysis was based on the fact that the CGRH depends on the propagation of the VHE photons across large distances, which is dependent on cosmology. \citet{barrau08} derive a lower limit of the Hubble constant, $H_{0}>74$~km~s$^{-1}$~Mpc$^{-1}$ at a 68\% confidence level, from the observation of $\gamma$-ray photons coming from a flare of the blazar Mrk~501, which was detected by HEGRA \citep{aharonian99}.

As described in \S\ref{sec:optdepth}, the cosmological dependence of the optical depth arises both from the cosmic volume containing the EBL (volume-redshift) and the distance propagated by the gamma-ray in the absorbing medium (distance-redshift). The knowledge of the EBL has improved dramatically in the last decade \citep[\eg][]{dwek13}. Recently, direct measurements in optical wavelengths of the EBL in the local universe \citep{matsuoka11,mattila17a,mattila17b} have confirmed previous indications \citep[\eg][]{aharonian06} of an EBL spectral intensity close to the estimations from integrated galaxy counts \citep[\eg][]{madau00,keenan10}. Furthermore, EBL models based on large multiwavelength galaxy data such as the ones by \citet{dominguez11a}, \citet{helgason12}, \citet{khaire15}, \citet{stecker16}, \citet{driver16}, \citet{franceschini17}, \citet{andrews18} and a better theoretical understanding of galaxy evolution \citep[\eg][]{somerville12,gilmore12} have allowed both the evaluation of the EBL at wavelengths where the detection is not possible yet and the convergence of different methodologies. Galaxy survey data can be used to construct the proper density of the absorbing EBL as a function of redshift. Together with the progress in the detection of $\gamma$-ray sources, this resulted in the first $H_{0}$ measurement using $\gamma$-ray attenuation, $H_{0}=71.8_{-5.6}^{+4.6}({\rm stat})_{-13.8}^{+7.2}({\rm syst})$~km~s$^{-1}$~Mpc$^{-1}$, by \citet{dominguez13b}. This measurement was followed by \citet{biteau15}, who used different $\gamma$-ray observations but reached compatible results, $H_{0}=88\pm 13 ({\rm stat}) \pm 13 ({\rm syst})$~km~s$^{-1}$~Mpc$^{-1}$.

\section{Methodology}

\subsection{General considerations}

We base our estimation of the Hubble constant on the hypothesis that the evolving EBL is sufficiently well described by the latest EBL models. In order to estimate the systematic uncertainties introduced by the EBL model selection, we use two models. First, the physically motivated model by \citet[][hereafter F10, see \S\ref{subsec:finke10}]{finke10} and second the observational model by \citet[][hereafter D11, see \S\ref{subsec:dominguez11}]{dominguez11a}. According to \citet{ebl18}, the F10 model is in excellent agreement with the LAT blazar $\gamma$-ray data, characterized by a significance of rejection of the model of only $0.4\sigma$. The D11 model is still compatible with the data, but was found to have a rejection significance of $2.9\sigma$ (which we do not consider significant), mainly due to discrepancies at high redshifts. These reported tensions between EBL models and $\gamma$-ray observations at high redshift may have some impact on our measurement of $\Omega_m$, but we expect negligible effect on the estimate of the Hubble constant (see \S\ref{sec:gra} for more details). These two models should provide a spread of the EBL model systematic uncertainties that are consistent with the $\gamma$-ray data. Therefore, the combination of these two models can provide an estimate of the systematic uncertainty introduced in the cosmological constraints by our uncertainty in the EBL knowledge. When combining constraints based on the two EBL models, we assume that both models are equally probable.

Optical depths as a function of energy and redshift are independently derived following both the F10 and D11 methodology. A 2D grid of values of the Hubble constant and matter density are fit to the optical depth data using the Monte Carlo Markov Chain method for sampling the posterior probability distribution. We assume a flat $\Lambda$CDM cosmology (which implies, $\Omega_{\Lambda}=1-\Omega_{m}$), well justified by all observations including the CMB. We set the uniform prior that $40\leq H_{0}\leq 95$~km~s$^{-1}$~Mpc$^{-1}$ and $0.05\leq \Omega_{m}\leq 0.6$ in agreement with other independent observational constrains. Then, a global fit is performed to the optical depths as a function of energy in fourteen redshift bins, twelve from \citet{ebl18}, and two from \citet{desai19}, leaving $H_{0}$ and $\Omega_{m}$ free. The data contain upper and lower limits that are considered in our fits by using the probability distributions from which each data point was derived.


Both the $\gamma$-ray measurements and the EBL models may be subject to hidden systematic uncertainties. We account for these effects by fitting a systematic error in $\delta\tau/\tau$ as an additional nuisance parameter. We assume that the systematic error is independent of the statistical uncertainties in the measurements of $\gamma$-ray attenuation. In order to exhaust all possible trends, we also assume that the systematic error is a power-law function of $\gamma$-ray energy and a scale factor $a=1/(1+z)$, where power-law indices are additional free parameters fitted to the data. The latter scaling accounts for possible larger systematic uncertainties at the highest redshift data, as expected from the aforementioned minimal, but persistent, discrepancies between the EBL estimates calculated for the standard cosmological model and the measured optical depths \citep[see Fig. 1 in][]{ebl18}. All cosmological constraints presented in our work are marginalized over the nuisance parameters describing the systematic errors. For both EBL models, the inferred systematic errors are subdominant (with the mean approximately of 0.03 whereas the statistical errors in the data are approximately 0.3). The models yield fully satisfactory fits with the reduced $\chi^{2}$ of 0.67 and 0.81 for F10 and D11, respectively.

\subsection{Optical depth dependence on cosmology}\label{sec:optdepth}
Pair production interactions between $\gamma$-ray and EBL photon produce an $\gamma$-ray optical depth $\tau$ that is analytically given by

\begin{equation}
\label{attenu}
\tau(E,z)=\int_{0}^{z} \Big(\frac{dl}{dz'}\Big) dz' \int_{0}^{2}d\mu \frac{\mu}{2}\int_{\varepsilon_{th}}^{\infty} d\varepsilon{'}\ \sigma_{\gamma\gamma}(\beta{'})n(\varepsilon{'},z'),
\end{equation}

\noindent where $E$ is observed energy and $z$ the redshift of the $\gamma$-ray source.\\

The energy threshold of the pair production interaction is given by the lower limit of the energy integral $\varepsilon_{th}$,

\begin{equation}
\label{threshold}
\varepsilon_{th}\equiv \frac{2m_{e}^2c^{4}}{E'\mu},
\end{equation}
\noindent where $E'$ is the energy of the $\gamma$ photon and $\varepsilon{'}$ the energy of the EBL photon (both in the rest-frame at redshift $z'$), and $\mu=(1-\cos \theta)$, with $\theta$ the angle of the interaction. The constant $m_{e}$ is the electron mass and $c$ is the vacuum speed of light.

In Equation~\ref{attenu}, the factor $n(\varepsilon{'},z')$ is the proper number density per unit energy of EBL photons, $\sigma_{\gamma\gamma}$ is the photon$-$photon pair production cross section, and $\beta{'}$ is

\begin{equation}
\label{beta}
\beta^{'}=\frac{\varepsilon_{th}}{\varepsilon{'}(1+z')^{2}}.
\end{equation}

Equation~\ref{attenu} shows a factor $dl/dz'=c|dt/dz'|$, which defines how the infinitesimal space element varies with redshift. For the standard Friedmann-Lema{\^i}tre-Robertson-Walker metric

\begin{equation} 
\label{peebles}
\Big|\frac{dt}{dz'}\Big|=\frac{1}{H_{0}(1+z')E(z')}
\end{equation}
\noindent with

\begin{equation}
\label{eq:Ez}
E(z') \equiv \sqrt{\Omega_{m}(1+z')^3+\Omega_{\Lambda}} \, \, , 
\end{equation}

\noindent and $H_{0}$, $\Omega_{m}$ and $\Omega_{\Lambda}$ are the parameters of the flat $\Lambda$CDM cosmology \citep{peebles93}.\\

From Equation~\ref{attenu}, we see that $\tau$ is dependent on cosmological parameters by two factors. First, the dependence given by the EBL density evolution $n(\varepsilon,z)$. Second, the dependence with the extragalactic $\gamma$-ray propagation through the Universe given by the factor $dl/dz$.

\subsection{Extragalactic background light photon evolution}\label{subsec:n}
Here, we calculate $n(\varepsilon,z)$ as a function of $H_{0}$ and $\Omega_{m}$ using two different independent methodologies. Initially, both EBL models were built adopting a standard $\Lambda$CDM cosmology where $H_{0}\p=70$~km~s$^{-1}$~Mpc$^{-1}$ and $\Omega\p_m=0.3$, for a flat Universe. This choice is compatible with the latest constraints from the Planck observations of the CMB, $\Omega_{m}=0.315\pm 0.007$ \citep{aghanim18}.

\subsubsection{Finke et al. (2010) model} \label{subsec:finke10}
The EBL is modeled by F10 in the following manner.  They first computed the stellar luminosity density, 
\begin{flalign}
\e j^{\rm stars}(\e; z) = m_e c^2 f_{\rm esc}(\e) \int dm \xi(m) 
\nonumber \\ \times
\int dz_1 \left| \frac{dt_\star}{dz_1} \right| \psi(z_1) \dot{N}_\star(\e; m, t_\star(z,z_1))\ ,
\end{flalign}
where $\e$ is the photon energy in $m_e c^2$ units, $m$ is the stellar
mass in $M_\odot$ units, $f_{\rm esc}(\e)$ is the fraction of photons
that escape dust absorption taken from \citet{driver08} and
\citet{razzaque09}, $\xi(m)$ is the initial mass function,
$\psi(z_1)$ is the star formation rate, and $\dot{N}_\star(\e; m,
t_\star(z,z_1))$ is the number of photons produced by stars as a
function of age and stellar mass \citep[see F10 and][for
  details]{eggleton89}.  Once $\e j^{\rm stars}(\e; z)$ was calculated,
  the luminosity density of the dust component $j^{\rm dust}(\e;
  z)$ was calculated self-consistently, assuming the photons absorbed
  by dust were reemitted in three dust components.  F10 found that
  $\xi$ from \citet{baldry03} and $\psi(z)$ using the \citet{cole01}
  parameterization, fit by \citet{hopkins06}, provided a good
  representation of the luminosity density data available at the time.
  Once the luminosity density from stars and dust are computed, the
  EBL photon density is calculated,

\begin{flalign}
n(\e; z) = \frac{1}{m_e c^2\e}\int_z^{z_{\rm max}} dz_1\ [j^{\rm stars}(\ep; z_1) + 
j^{\rm dust}(\ep; z_1)] \left| \frac{dt_\star}{dz_1} \right|
\end{flalign}
where $\ep = (1+z_1)\e$.  From this, the absorption optical depth is
computed (Equation \ref{attenu}).

This model depends on the cosmological parameters ($H_{0}$ and $\Omega_m$) 
through $|dt/dz|$ as described in \S\ref{sec:optdepth} and through
$\psi(z)$.  \citet{hopkins06} fit a function $\psi_{HB06}(z)$ to a variety of 
star formation rate data assuming a flat universe with $H_{0}=70$~km~s$^{-1}$~Mpc$^{-1}$ and
$\Omega_m=0.3$.  Therefore, we must also modify the star formation rate
for different cosmological values \citep[\eg][]{ascasibar02},
\begin{flalign}
\psi(z) = \psi_{HB06}(z)\frac{H_{0} E(z)}{H_{0}\p E\p(z)}\
\end{flalign}
where primed quantities are computed with fiducial cosmological parameters
$H_{0}\p=70$~km~s$^{-1}$~Mpc$^{-1}$ and $\Omega\p_m=0.3$.

\subsubsection{Dom\'inguez et al. (2011) model} \label{subsec:dominguez11}
The EBL was empirically derived by D11 from two main ingredients. First, the estimation of galaxy SED-type fractions based upon a multiwavelength catalog of around 6,000 galaxies drawn from the All-wavelength Extended Groth strip International Survey \citep[AEGIS, ][]{davis07}.

Second, the $K$-band galaxy luminosity functions (LFs) from \citet{cirasuolo10} were used to give the number of galaxies per unit volume and magnitude in the near-IR from the local Universe of to $z\sim4$. The galaxy LFs are described by Schechter functions \citep{schechter76}, whose parameters depend on cosmology. Schechter functions are parameterized by three quantities: $\phi_{0}(z)$, $M_{*}(z)$, and $\alpha$ (these are the normalization, a characteristic absolute magnitude, and the faint-end slope). It is then possible to compute the Schechter LFs for a new set of  $\Lambda$CDM cosmological parameters ($h$, $\Omega_{m}$, and $\Omega_{\Lambda}$) providing the values of $\phi^{'}_{0}(z)$ and $M^{'}_{*}(z)$ obtained adopting the fiducial parameters ($h'$, $\Omega_{m}^{'}$, and $\Omega_{\Lambda}^{'}$). We note that $h$ is the dimensionless parameter $h=H_{0}/100$, and provide the equations for converting the Schechter LF from a fiducial set of cosmological parameters to another choice (written in absolute magnitudes), \ie

\[
\phi(M,h,\Omega_{m},\Omega_{\lambda},z)=0.4\ln(10)\phi_{0} \times 10^{0.4(M_{*}-M)(\alpha+1)}
\]
\begin{equation}
\label{eq:last}
\hspace{0.5cm}\times \exp[-10^{0.4(M_{*}-M)}]  \hspace{1.6cm}\textrm{[Mpc$^{-3}$~Mag$^{-1}$]}
\end{equation}

\begin{equation}
\phi_{0}=\phi_{0}'\Big(\frac{h}{h'}\Big)^{3} \frac{E(z)}{E'(z)} \Big[\frac{F'(z)}{F(z)}\Big]^{2}
\end{equation}

\begin{equation}
M_{*}=M_{*}'+5\log_{10}\Big[\frac{H_{0}}{H_{0}'}\frac{F'(z)}{F(z)}\Big]
\end{equation}

\noindent where

\begin{equation} \label{eq:F}
F(z) = \int_{0}^{z}\frac{dz'}{E(z')},
\end{equation}

\noindent where $E(z')$ is given by Equation~\ref{eq:Ez}.

Therefore, from Equation~\ref{eq:last} and following the methodology described by D11, it is possible to calculate the luminosity densities, and thus the EBL photon density evolution $n(\varepsilon,z)$ for different cosmologies. 

\begin{figure*}
\includegraphics[width=6.45cm,trim={0 0 1.2cm 0},clip]{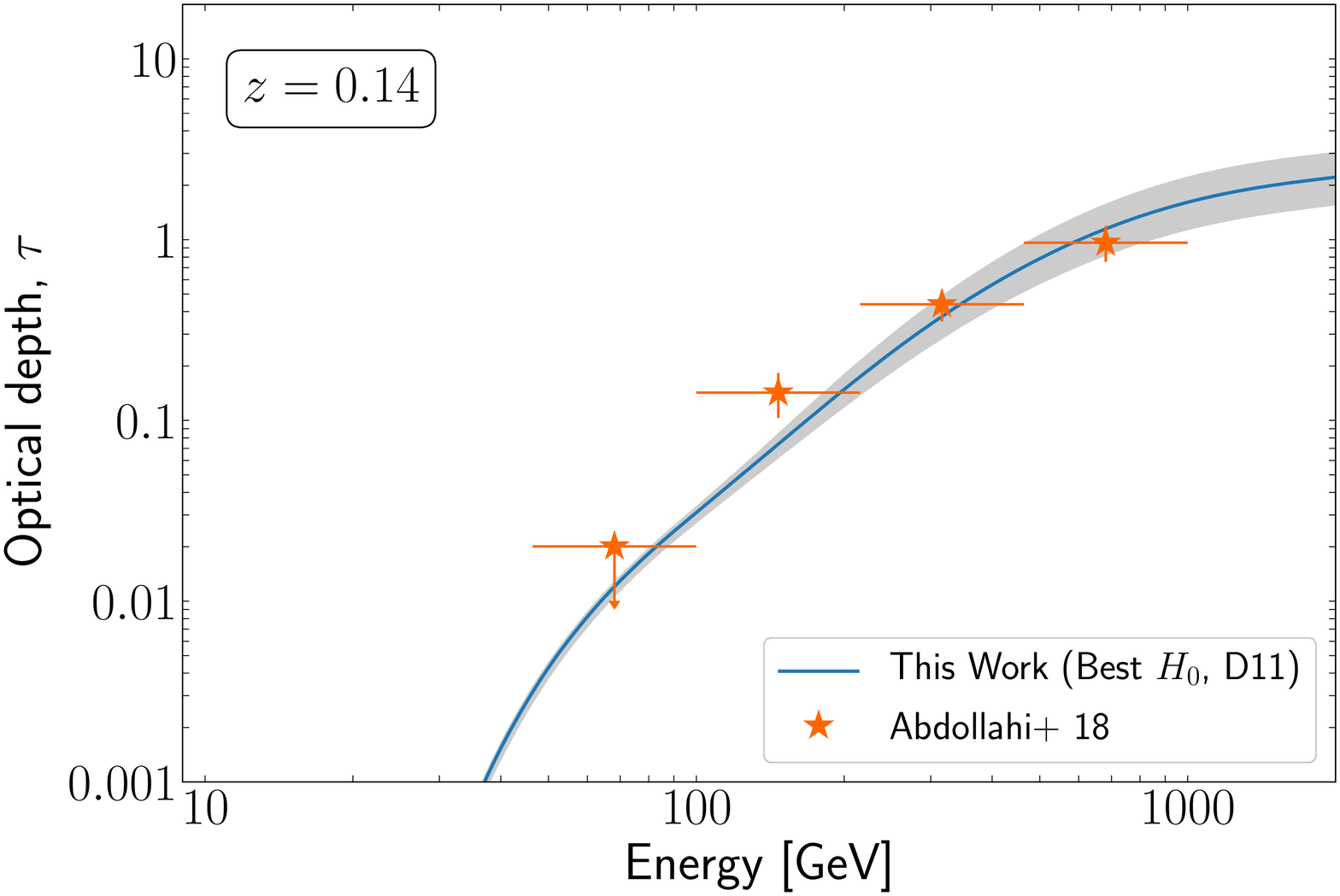}
\includegraphics[width=5.5cm,trim={4.2cm 0 1.2cm 0},clip]{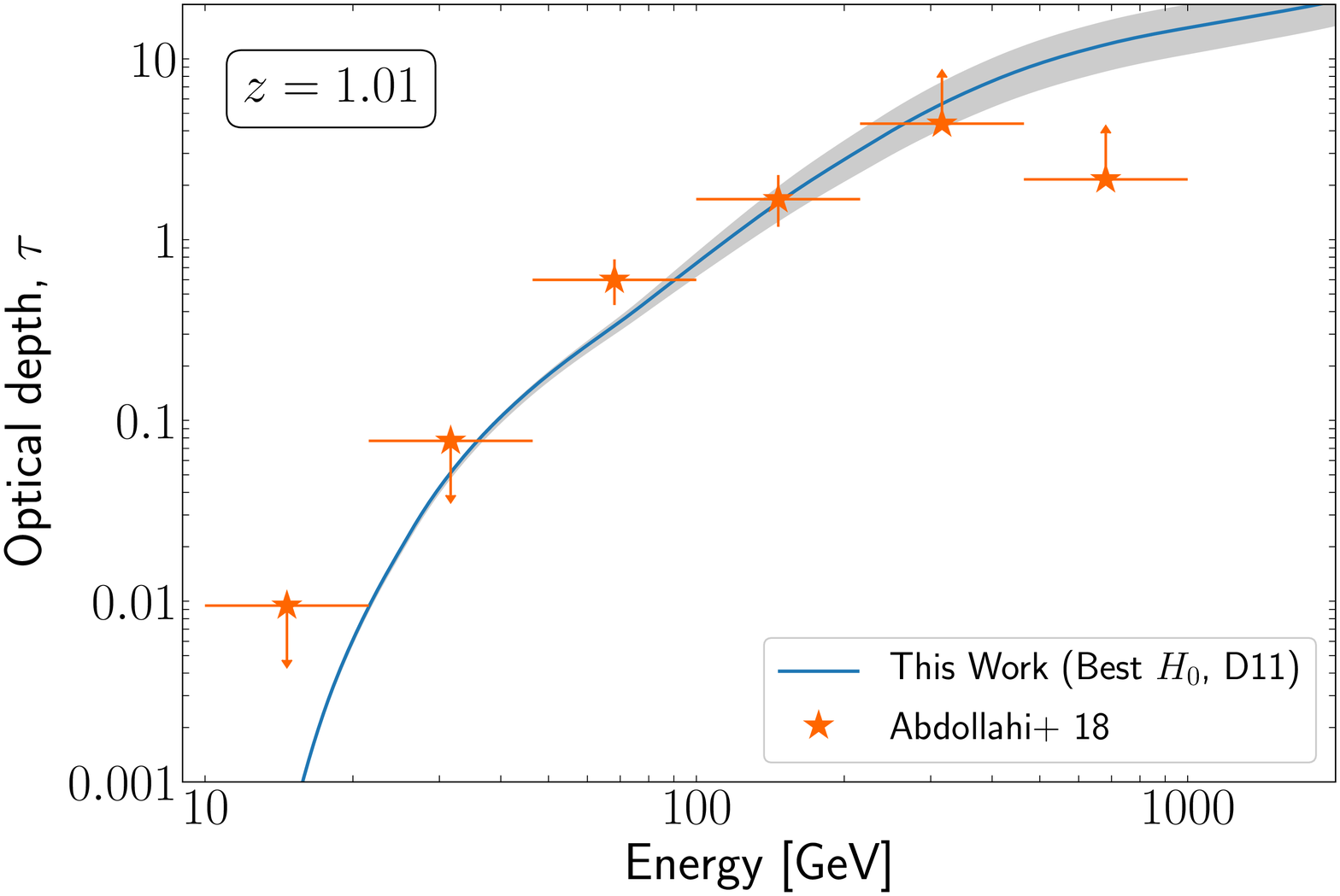}
\includegraphics[width=5.5cm,trim={4.2cm 0 1.2cm 0},clip]{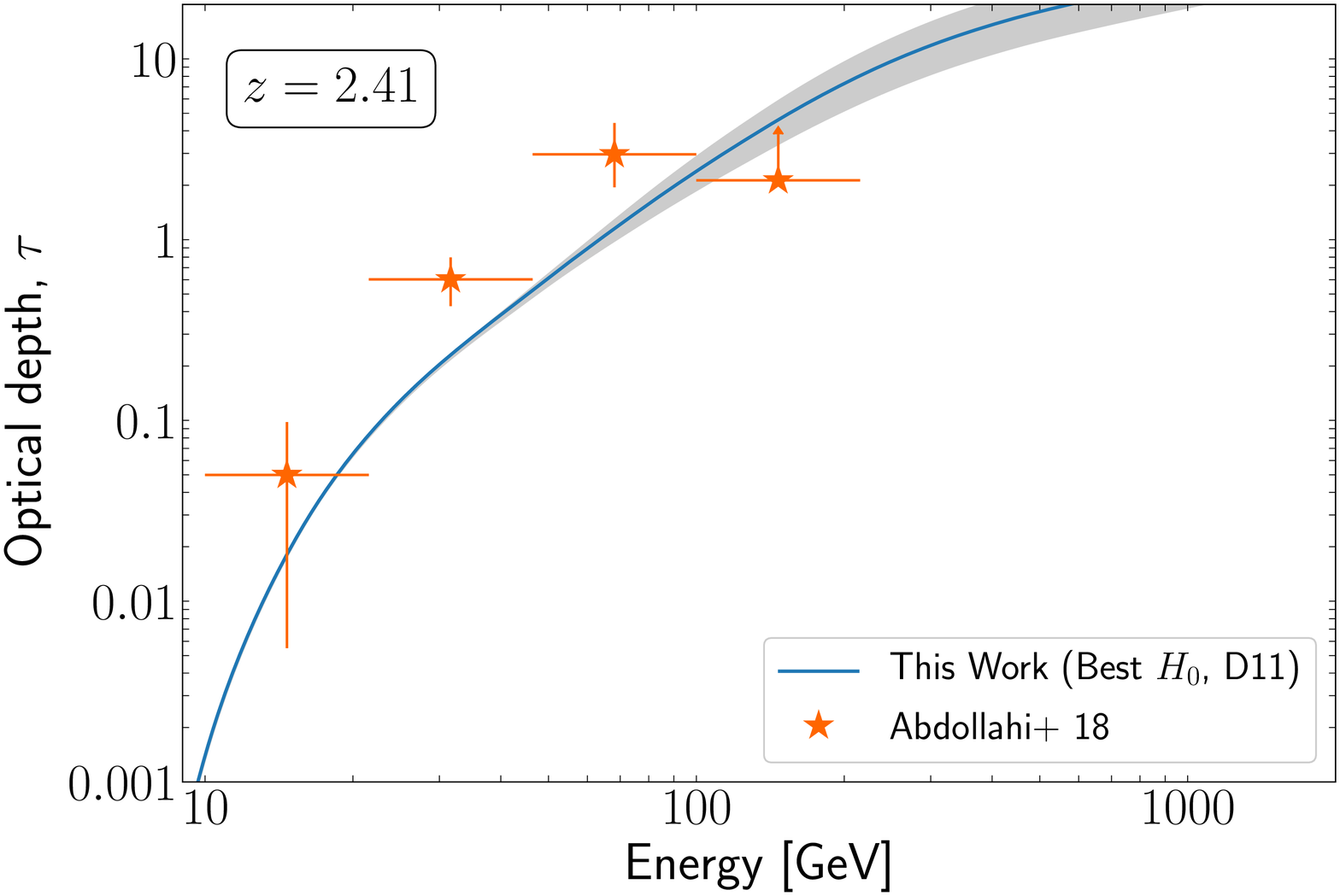}
\caption{Examples of the optical depth dependence on $H_{0}$ for three redshifts ($z=0.14$, 1.01, and 2.41). These plots are produced by fixing $\Omega_{m}=0.32$ and varying $H_{0}$, shown as a gray band, from 40~km~s$^{-1}$~Mpc$^{-1}$ (upper bound) to 95~km~s$^{-1}$~Mpc$^{-1}$ (lower bound) assuming the EBL model by D11. We see that the dependence of the optical depth with $H_{0}$ happens at energies larger than 100~GeV.}
\label{fig:opdepthH0}
\end{figure*}

\begin{figure}
\includegraphics[width=\columnwidth]{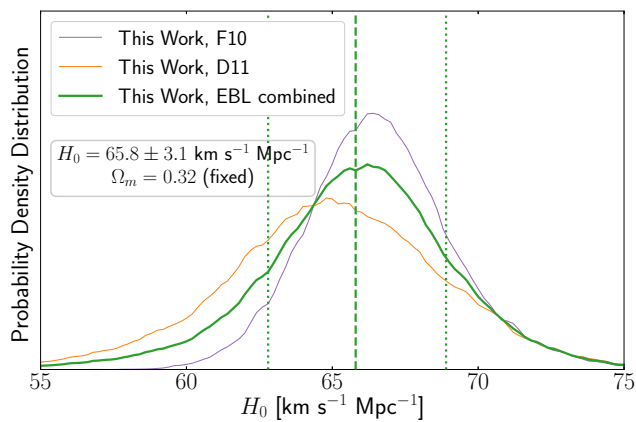}
\caption{Posterior probability distribution as a function of $H_{0}$ (fixed $\Omega_{m}=0.32$) for the F10 (purple) and D11 (orange) EBL models, and the combined results (green, see the text for details). The median of the combined results is also plotted with $1\sigma$ containment (green vertical lines).}
\label{fig:1D}
\end{figure}

\subsection{Optical depth data}
The optical depth data are taken from \citet{ebl18} and \citet[][see these references for details, in particular, Figure~2 of the latter one]{desai19}. For completeness, we will give here a brief overview.

In \citet{ebl18}, optical depths are estimated by measuring the $\gamma$-ray attenuation from a sample of 739 blazars plus one $\gamma$-ray burst, all detected by {\it Fermi}-LAT. These optical depths are given in twelve redshifts bins reaching $z\sim 3.10$. These redshift bins are chosen in such a way that the signal's strength is the same in each one of them. The optical depths are given in six logarithmically equally spaced energy bins from approximately 10~GeV up to 1000~GeV. The \citet{ebl18} results are especially relevant for constraining $\Omega_{m}$ because the larger dependence of the optical depth with $\Omega_{m}$ occurs at the larger redshifts.


\citet{desai19} use a sample of 38 blazars detected by IACTs leading to a measurement of optical depths in two redshift bins up to $z\sim 0.6$. These optical depths are measured in four equally spaced logarithmic energy bins from 0.1~TeV up to approximately 20~TeV. These results from \citet{desai19} are especially important for measuring $H_{0}$ because the largest dependence of the optical depth with $H_{0}$ occurs at the higher energies and lower redshifts.

\begin{figure*}
\includegraphics[width=6.45cm,trim={0 0 1.2cm 0},clip]{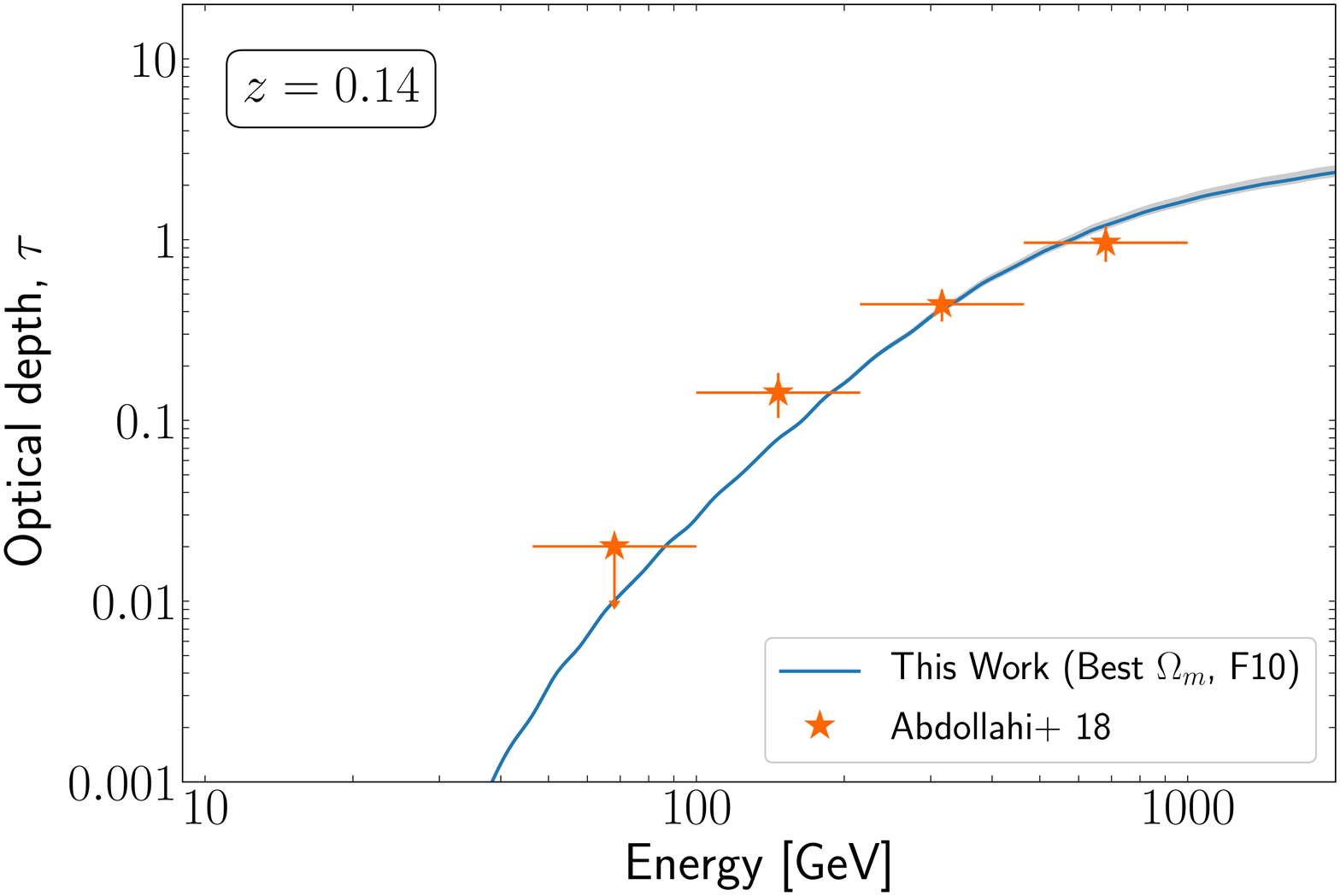}
\includegraphics[width=5.5cm,trim={4.2cm 0 1.2cm 0},clip]{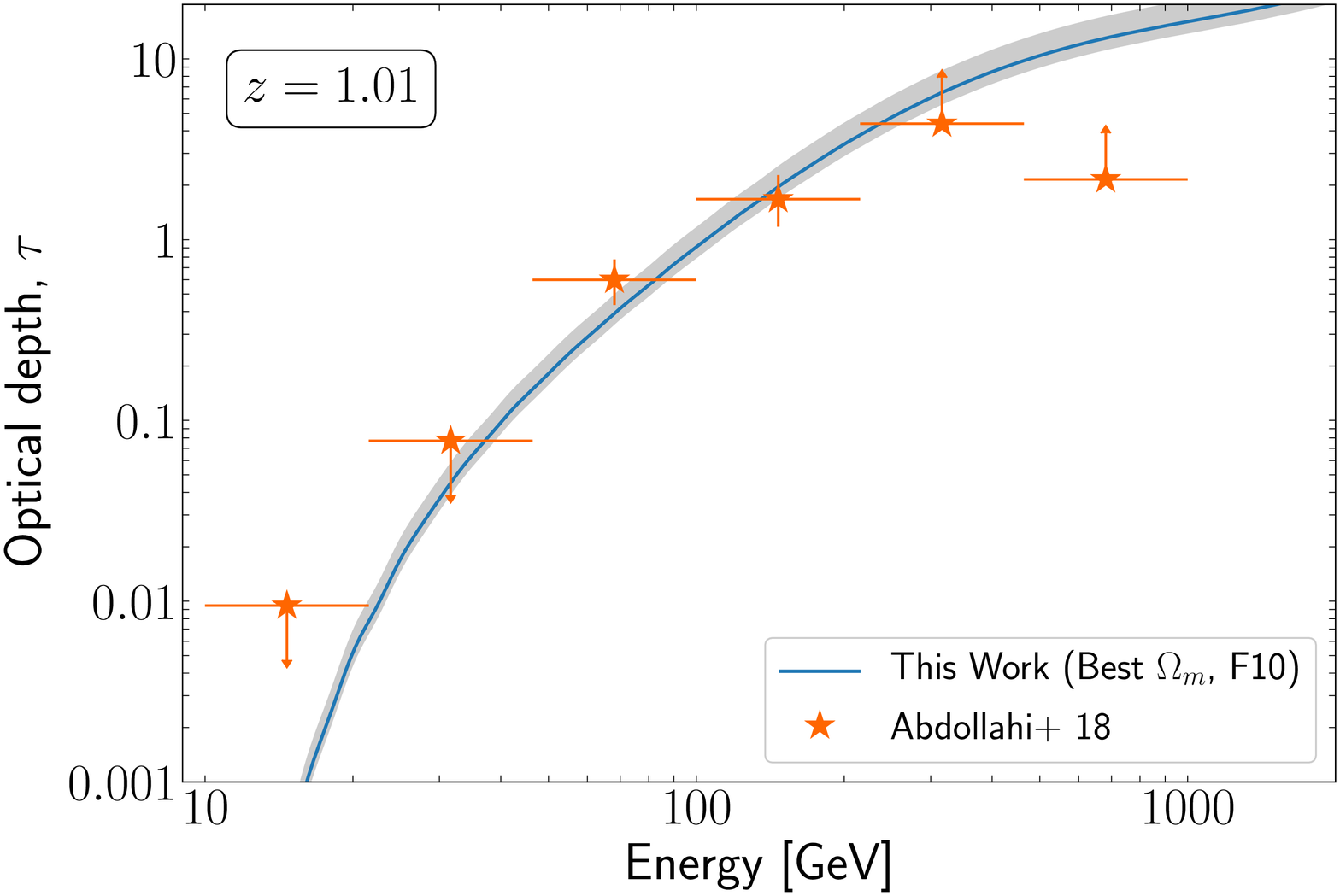}
\includegraphics[width=5.5cm,trim={4.2cm 0 1.2cm 0},clip]{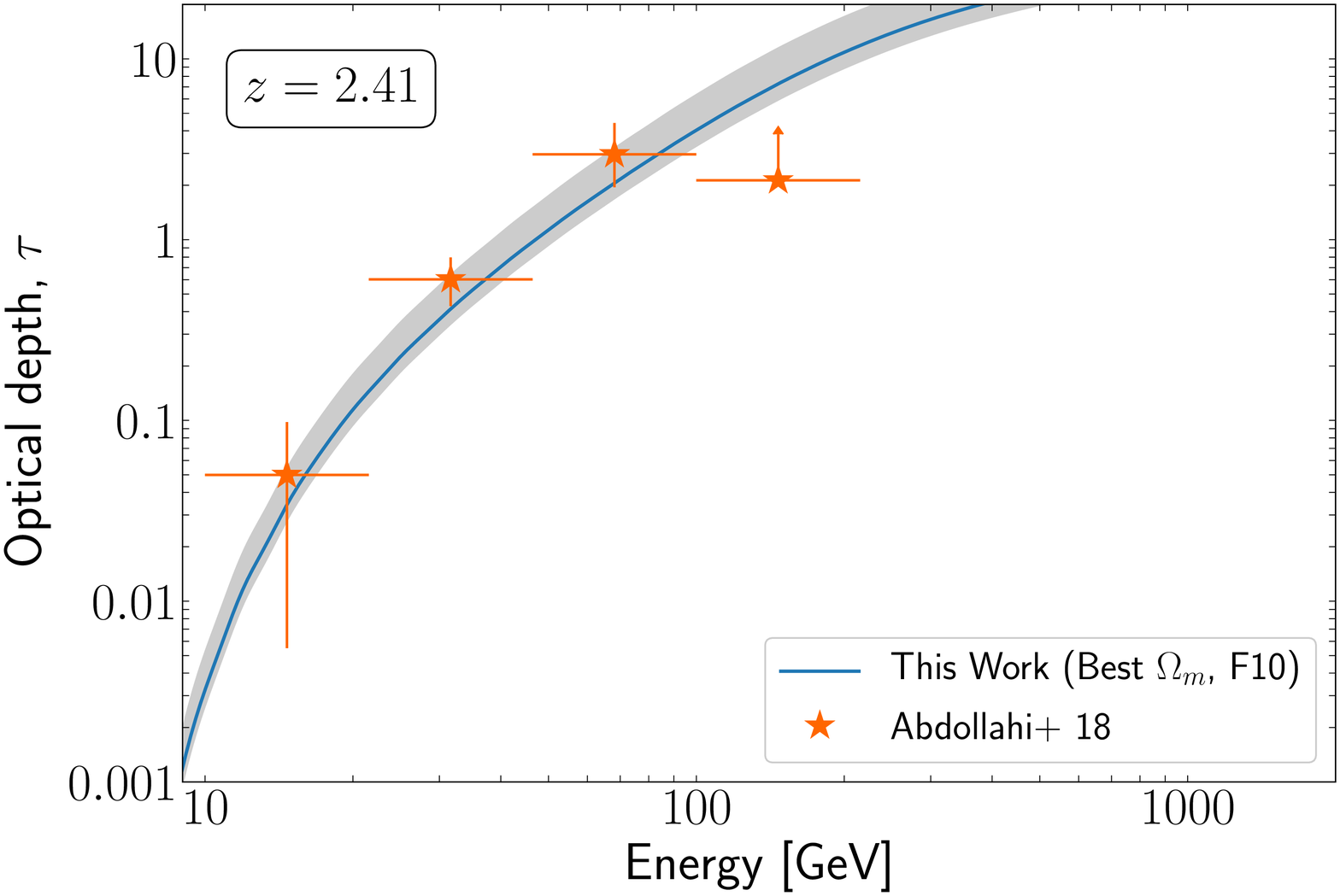}
\caption{Examples of the optical depth dependence on $\Omega_{m}$ for three redshifts ($z=0.14$, 1.01, and 2.41). These plots are produced by fixing $H_{0}$ to the most likely value and varying $\Omega_{m}$, shown as a gray band, from 0.05 (upper bound) to 0.6 (lower bound) assuming the EBL model by F10. We see the little dependence of the optical depth with $\Omega_{m}$ and that the variation is larger for the higher redshifts.}
\label{fig:opdepthWM}
\end{figure*}

\begin{figure}
\includegraphics[width=\columnwidth]{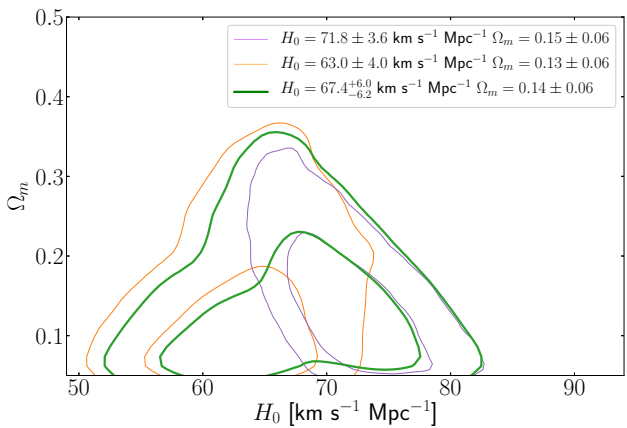}
\caption{Posterior probability ($1\sigma$ and $2\sigma$) contours for $H_{0}$ and $\Omega_{m}$ assuming the EBL model by F10 (purple) and D11 (orange), and also combining the results from both models (green). We consider both models to be equally likely.}
\label{fig:likelihood}
\end{figure}


\section{Results}
In this section, we present the results of applying our maximum likelihood methodology to $\gamma$-ray attenuation results. We also combine them with other strategies for deriving cosmological parameters from the literature.

\subsection{Gamma-ray attenuation}\label{sec:gra}
As a first exercise, we fix $\Omega_{m}=0.32$ to match the Planck cosmology \citep{planck19}, and proceed with fitting only $H_{0}$. This strategy will show how our better data improve the measurement of $H_{0}$ relative to the previous work by \citet{dominguez13b}. For illustration purposes, we see in Figure~\ref{fig:opdepthH0} how the effect works. The $H_{0}$ dependence occurs at energies larger than about 100~GeV and increases with energy. Since the optical depth increases with redshift and becomes more difficult to measure, the strongest constraints on $H_{0}$ come from the lower redshifts. Figure~\ref{fig:1D} shows our independent and combined probability density distribution using the F10 and D11 EBL models. We obtain $H_{0}=65.8_{-3.0}^{+3.1}$~km~s$^{-1}$~Mpc$^{-1}$, an estimate that includes statistical plus the systematic uncertainty from the EBL models. These uncertainties are at the 5\% level. Fixing $H_{0}=68$~km~s$^{-1}$~Mpc$^{-1}$ and searching for the most likely matter density value leads to $\Omega_{m}=0.17_{-0.08}^{+0.07}$.

Now, we extend our parameter space exploring simultaneously the grid of $H_{0}$ and $\Omega_{m}$. We show in Figure~\ref{fig:opdepthWM} the effect of varying $\Omega_{m}$. We can see that in general the optical depths are not strongly dependent on $\Omega_{m}$. In particular, the optical depth barely depends on $\Omega_{m}$ for the lowest redshift, but the dependence becomes stronger for the larger redshifts. Figure~\ref{fig:likelihood} shows the resulting likelihood contours. The most likely values from the $\gamma$-ray methodology are $H_{0}=67.4_{-6.2}^{+6.0}$~km~s$^{-1}$~Mpc$^{-1}$ and $\Omega_{m}=0.14_{-0.07}^{+0.06}$. The Hubble constant is measured with a relative error of 9\%, whereas the matter density parameter is measured with a relative error of 50\%. We note that these fit parameters are posterior probability means. 

\begin{figure}
\includegraphics[width=\columnwidth]{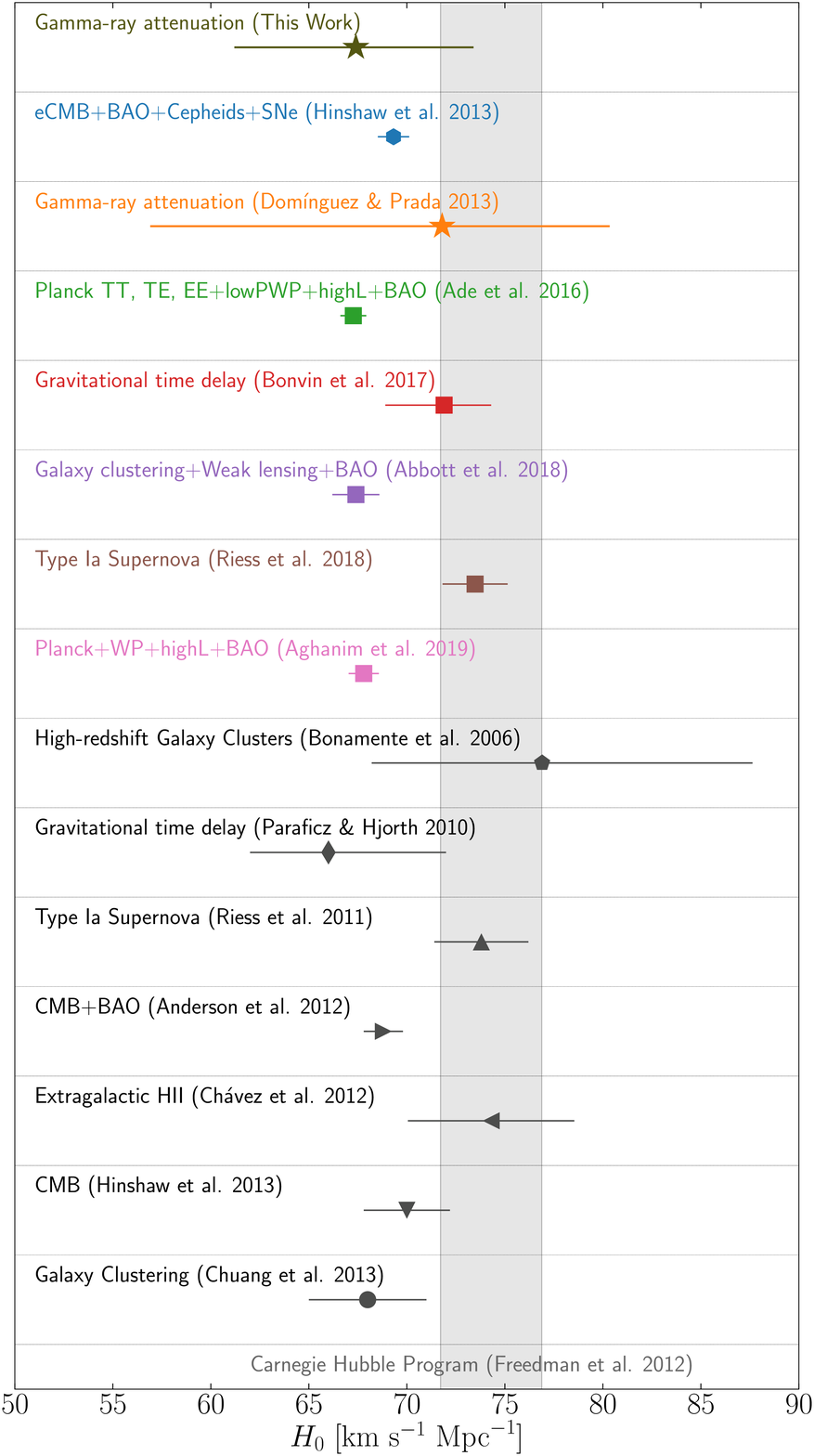}
\caption{Comparison of $H_{0}$ from different methodologies. The measurement from the Carnegie Hubble Program \citep{freedman12} is shown as a gray rectangle for easier comparison with other results. Other results are from \citet{bonamente06}, \citet{paraficz10}, \citet{riess11}, \citet{chavez12}, \citet{anderson12}, \citet{suyu12}, \citet{hinshaw13}, \citet{chuang13}, \citet{dominguez13b}, \citet{ade16}, \citet{bonvin17}, \citet{DESH02018}, \citet{riess18a} and \citet{planck19}.}
\label{fig:comparison}
\end{figure}

In Figure~\ref{fig:comparison}, we compare $H_{0}$ estimated from different methodologies in the literature.

\subsection{Combination of results from different probes}

Here we combine our cosmological constraints from the attenuation of $\gamma$-rays with results from the primary cosmological probes. In particular, 
we consider a compilation of the baryon acoustic oscillation (BAO) observations and Type Ia supernovae. Our BAO sample includes the angular 
diameter distances and Hubble parameters from the SDSS-III/BOSS \citep{Ala2015} measured at redshifts $z=0.38$, 0.5, and 0.61 using prereconstruction 
(independent of cosmological model) methods \citep{Ala2017}, distance measurements from the 6dF survey at $z=0.106$ \citep{Beu2011} and from 
the Main Galaxy Sample of the SDSS (SDSS-MGS) at $z=0.15$ \citep{Ros2015}. For the SN data, we use distance moduli from a joint likelihood analysis of SDSS-II and SNLS type Ia supernovae samples \citep{Bet2014}.

\begin{figure}
\includegraphics[width=\columnwidth]{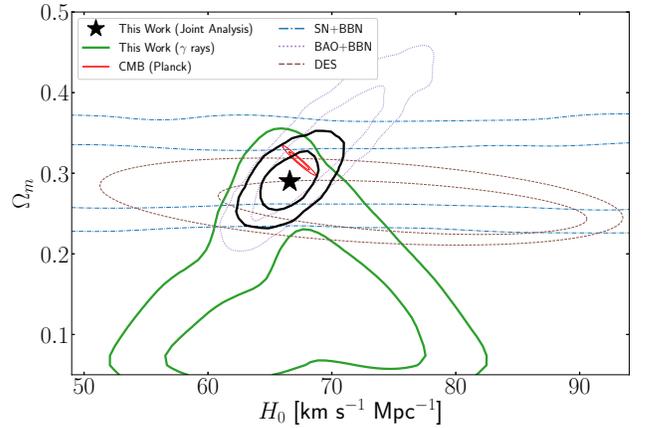}
\caption{Measurements of the Hubble constant and matter density (1$\sigma$ and 2$\sigma$) using $\gamma$-ray attenuation (green), supernovae plus Big Bang nucleosynthesis (SN+BBN, blue), baryonic acoustic oscillations plus Big Bang nucleosynthesis (BAO+BBN, purple), clustering and weak lensing data (DES, brown), the cosmic microwave background (Planck, red) and a joint likelihood of BAO+BBN+SN+$\gamma$ (black). The maximum likelihood value is at $H_{0}=66.6\pm 1.6$~km~s$^{-1}$~Mpc$^{-1}$ and $\Omega_{m}=0.29\pm 0.02$ (black star).}
\label{fig:all}
\end{figure}

\begin{deluxetable}{lcc}
\tablehead{
\colhead{Methodology} &
\colhead{$H_{0}$~[km~s$^{-1}$~Mpc$^{-1}$]} &
\colhead{$\Omega_{m}$}
}
\startdata
Gamma-ray Attenuation & $65.8_{-3.0}^{+3.1}$ & 0.32 (fixed)\\
Gamma-ray Attenuation & 68 (fixed) & $0.17_{-0.08}^{+0.07}$ \\
Gamma-ray Attenuation & $67.4_{-6.2}^{+6.0}$ & $0.14_{-0.07}^{+0.06}$ \\
Joint Likelihood Analysis & $66.6\pm 1.6$ & $0.29\pm 0.02$\\
\enddata
\caption{The favored values of $H_{0}$ and $\Omega_{m}$ from $\gamma$-ray attenuation (fixing $\Omega_{m}$, $H_{0}$, and also leaving free both parameters) and from our joint analysis of BAO+BBN+SN+$\gamma$ results. Uncertainties are given at 1$\sigma$.}
\end{deluxetable}

The BAO signal depends primarily on $\Omega_{m}$, $H_{0}$ and the baryon density parameter $\Omega_{\rm b}h^{2}$. Therefore, BAO observations 
at a single redshift result in strong degeneracies between these three parameters. Following \citet{addison13}, \citet{aubourg15}, and \citet{Lin2017}, we break this degeneracy by assuming a prior on $\Omega_{\rm b}h^{2}$ from the Big Bang nucleosynthesis theory constrained by the local measurements of the primordial light element abundances. In particular, we adopt $100\Omega_{\rm b}h^{2}=2.208\pm0.052$ from the recent precise measurement of the primordial deuterium abundance in the most metal-poor damped Lyman-$\alpha$ system \citep{Cooke2016}. We also use the COBE/FIRAS measurement of the temperature of the cosmic microwave background radiation, i.e. $T_{\rm CMB}=(2.7255\pm0.0006$)~K \citep{Fixsen2009}. This makes the parameter set complete for calculating the sound horizon scale. For this part of the computation we use the camb code \citep{Lew1999}.

Figure~\ref{fig:all} shows marginalized constraints on $\Omega_{m}$ and $H_{0}$ from the BAO data with the BBN prior. We emphasize that 
these results are independent of any cosmological constrain from the CMB. In Figure~\ref{fig:all} is also shown analogous constraints from the supernova data. The lower bound in $\Omega_{m}$ results from adopting the same BBN prior which naturally requires $\Omega_{\rm b}\leq \Omega_{m}$.

Our constraints on $\Omega_{m}$ and $H_{0}$ are fully consistent with both BAO and SN observations. Figure~\ref{fig:all} also demonstrates the potential of gaining precision in the Hubble constant determination by combining our $\gamma$-ray measurement with the BAO+BBN constraints. From a joint likelihood analysis of the data sets we find $H_{0}=66.4_{-1.9}^{+1.8}$~km~s$^{-1}$~Mpc$^{-1}$ and $\Omega_{m}=0.29\pm 0.04$. Adding SN data has a marginal effect on the final constraints and results in $H_{0}=66.6\pm 1.6$~km~s$^{-1}$~Mpc$^{-1}$ and $\Omega_{m}=0.29\pm 0.02$. Figure~\ref{fig:all} shows the corresponding marginalized constraints. For the sake of comparison with other existing measurements, we also plot constraints from the Planck observations of the CMB \citep{planck19}, including the temperature, polarization, and lensing data, and also from a cosmological inference that combines clustering and weak lensing data from the first year of observations by the Dark Energy Survey \citep{DESH02018}.

\section{Discussion and conclusions} \label{sec:summary}
Our methodology based on comparing $\gamma$-ray attenuation data with estimates from EBL models leads to a measurement of $H_{0}=65.8_{-3.0}^{+3.1}$~km~s$^{-1}$~Mpc$^{-1}$ (this is a relative error of 5\%), when $\Omega_{m}=0.32$ is fixed. When $\Omega_{m}$ is also left free, we find $H_{0}=67.4_{-6.2}^{+6.0}$~km~s$^{-1}$~Mpc$^{-1}$ and $\Omega_{m}=0.14_{-0.07}^{+0.06}$, including a detailed analysis of systematic uncertainties (considering also those introduced by two state-of-the-art EBL models).

We stress that our analysis is a significant step forward relative to previous cosmological measurements using $\gamma$-ray attenuation \citep{dominguez13b,biteau15}. First, the previous works are based on more limited energy data. In particular, the former work uses only the information provided by the CGRH, that is, a measurement of the optical depth at a single energy, whereas we take advantage of optical depth data as a function of energy. Second, they use blazar data only at low redshift $z\le 0.6$; however, in the present analysis we cover approximately the range $0.02\le z \le 3$. These improvements in the data allow us to simultaneously explore the values of $H_{0}$ and $\Omega_{m}$. Third, this analysis also presents for the first time an analysis of some systematic biases from using this methodology, including an estimate of the uncertainty introduced by two EBL models. Fourth, we have combined the $\gamma$-ray attenuation results in a joint likelihood analysis with other independent, complementary, and more mature techniques.

Our measurements support a value of $H_{0}$ that is closer to that one found by the BAO methodology rather than the higher value from the Cepheids. Interestingly, the $H_{0}-\Omega_{m}$ contours from $\gamma$-ray attenuation are roughly orthogonal to results from other techniques, which makes our results nicely complementary to those from other probes. In order to improve the $H_{0}$ measurement we need to measure optical depths up to the largest possible energies. This is difficult with LAT because of the limited photon statistics. However, it may be possible with the future Cherenkov Telescope Array \citep[CTA,][]{hinton19}.

These results illustrate the increasing potential of using $\gamma$-ray observations to constrain cosmology. In particular, our analysis paves the way for future cosmological measurements using $\gamma$-ray data from blazars and $\gamma$-ray bursts detected with {\it Fermi}-LAT, the current generation of IACTs, and also the upcoming CTA.

\section*{Acknowledgments}
The authors thank Javier Coronado-Bl{\'a}zquez, Jose Luis Contreras, and the anonymous referee for helpful comments. AD is thankful for the support of the Ram{\'o}n y Cajal program from the Spanish MINECO. RW was supported by a grant from VILLUM FONDEN (Project No. 16599). JDF is supported by NASA under contract S-15633Y. MA  and A.~Desai acknowledge funding support from NSF through grant AST-1715256. KH acknowledges support from the Icelandic Research Fund, grant number 173728-051. DH and LM acknowledge funding support from NASA through grant 80NSSC18K1720.

\bibliographystyle{apj}
\bibliography{biblio}

\label{lastpage}
\end{document}